\documentclass[%
 reprint,
amsmath,amssymb,
aps,
nofootinbib,
]{revtex4-2}

\usepackage{graphicx}% Include figure files
\usepackage{dcolumn}% Align table columns on decimal point
\usepackage{bm}% bold math
\usepackage{physics}
\usepackage{float}
\usepackage{verbatim}
\usepackage[small,tight,FIGTOPCAP]{subfigure}
\usepackage{amsmath}
\usepackage{amsfonts}
\usepackage{amssymb}
\usepackage[]{mathrsfs}
\usepackage{xcolor}
\usepackage{bbm}
\usepackage [english]{babel}
\usepackage{graphicx}% Include figure files
\usepackage{dcolumn}% Align table columns on decimal point
\usepackage{bm}% bold math
\usepackage{physics}
\usepackage{bbm}
\usepackage{appendix}
\usepackage{hyperref}% add hypertext capabilities
\usepackage{color}
\usepackage{hyperref}
\usepackage{cancel}
\newcommand\identity{1\kern-0.25em\text{l}}
 
\begin{document}

%\linenumbers
\preprint{APS/123-QED}

%\title{Trapped-ion motional SU(2) and SU(1,1) interferometers with sensitivities near the Cram\'er-Rao bound}
%\title{Programmable continuous-variable quantum circuits using trapped ions with phase sensitivities near the Cram\'er-Rao bound}
%\title{SU(2) and SU(1,1) interferometry using two-mode parametric resonances of trapped ion motional states}
\title{Two-mode squeezing and SU(1,1) interferometry with trapped ions}

\author{J. Metzner$^{1}$, A. Quinn$^{1}$, S. Brudney$^1$}
\author{I.D. Moore$^{1}$}\thanks{Present address: IonQ Inc, 4505 Campus Dr, College Park, Maryland 20740, USA}
\author{S.C. Burd$^2$, D.J. Wineland $^1$}
\author{D.T.C. Allcock$^1$}\thanks{E-mail: dallcock@uoregon.edu}
\affiliation{$^1$Department of Physics, University of Oregon, Eugene, OR, USA}
\affiliation{$^2$Department of Physics, Stanford University, Stanford, CA, USA}

\date{\today}% It is always \today, today
\begin{abstract}

We experimentally implement circuits of one and two mode operations on two motional modes of a single trapped ion. This is achieved by implementing the required displacement, squeezing, two-mode squeezing, and beamsplitter operations using oscillating electric potentials applied to the trap electrodes. The resulting electric fields drive the modes resonantly or parametrically without the need for optical forces. As a demonstration, we implement SU(2) and SU(1,1) interferometers with phase sensitivities near the Cram\'er-Rao bound. We report a maximum sensitivity of a SU(2) interferometer within 0.67(5)\,dB of the standard quantum limit (SQL) as well as a single and two-mode SU(1,1) sensitivity of 5.9(2)\,dB and 4.5(2)\,dB below the SQL respectively. 

\end{abstract}

\maketitle

Quantum-mechanical harmonic oscillators, described by infinite dimensional, non-commuting observables, are currently being utilized for quantum sensing~\cite{Zhuang2020,Sha2022} and studies of continuous variable quantum computing (CVQC)~\cite{Lloyd1999}. The quantum-mechanical harmonic oscillator is realized in trapped-ions, nano-mechanical systems~\cite{Regal2008} and has an analog with microwave photons in superconducting circuits~\cite{Oliver2005} and optical photons in photonic systems~\cite{Aasi2013}. Recent work, using motional states of trapped-ions has shown the ability to detect small displacements~\cite{Burd2019,Gilmore2021} and encode logical qubits~\cite{Fluhmann2019,DeNeeve2022}using a single mode and to measure phase shifts in entangled two mode states~\cite{PhysRevLett.121.160502}.  The use of two-mode squeezed states could provide a quantum advantage~\cite{Park2023}(larger quantum signal to noise ratio with the same number of resources) for sensing and significant work has been done to generate these states in other platforms~\cite{Xue2007,Woolley2014,Leong2023,cao2023joint} with some work theoretical exploring how to generate these states in trapped-ions~\cite{Nikolova_2023,ZENG2002427,PhysRevA.71.063817}. Two-mode squeezed states have already proven useful for quantum sensing using SU(1,1) interferometers~\cite{Jing2011,Eberle2013,Ou2020,Wang2010} in photonics, but can be difficult to generate efficiently and suffer from photon loss. 

In the following, we demonstrate different classes of interferometers using the motional modes of a single trapped atomic ion. With a demonstration of these interferometers, we show the ability to concatenate several operations on the oscillator states of two modes in a phase coherent manner. Performing these operations with digitally programmable frequency sources for relative phase control simplifies significant calibration challenges when developing more complicated circuits.

Trapped ions are well suited for such demonstrations due to the harmonic nature of the trap and the potential to achieve relatively long mechanical oscillator coherence times~\cite{Myatt2000}. Reconstructing the motional states of these oscillators is achieved by coupling them to the ion's internal states via motional sideband transitions and measuring the internal state populations~\cite{Meekhof1996}. In the limit of a small number of motional quanta these transitions can approximate a phonon counting type measurement. This enables a homodyne type detection scheme which is used to quantify the sensitivity we achieve with the different classes of interferometers demonstrated below.

Yurke \emph{et al.}~\cite{Yurke1986} proposed schemes to reach the Heisenberg limit~\cite{Pezze2009} using different classes of interferometers. Here, we work within the framework of the Cram\'er-Rao bound~\cite{Agarwal2022,Zwierz2010} which sets the ultimate limit for the resolution of a sensing scheme. This bound is described by a parameter variance, instead of the Heisenberg limit which is set by the number of resources used, i.e. photon or phonon number~\cite{Zwierz2010}, making it more suitable to the limits of the experiments reported here. The required interactions for the interferometers (described below) are colloquially called displacing, beam splitting, single-mode squeezing, and two-mode squeezing. For trapped ions, these interactions can be implemented with laser beams~\cite{Leibfried2002} or in a laser-free method using additional potentials applied to the trap electrodes~\cite{Serafini2009,Heinzen1990}. Performance of these interferometers can, in practice, be limited only by the motional heating of the ion resulting from fluctuating patch potentials on the trap electrodes~\cite{Deslauriers2006,Hite2013}. Here, we demonstrate an SU(2) interferometer (Mach-Zender), which has previously been demonstrated in~\cite{Leibfried2002,Gorman2014,Ding2017}, (coherent input states are used here in contrast to these works), as well as single-mode and two-mode SU(1,1)\footnote{The Lie-group labelling is used to refer to the groups that contain the operators that generate these states.} interferometers. Each of the interferometers described here is sensitive to different phase shifts, which are single-mode phase $\phi_a$ (single-mode SU(1,1)), difference phase $\phi_a -\phi_b$ (SU(2)), or sum phase $\phi_a+\phi_b$ (two-mode SU(1,1)), where the $a,b$ labels refer to different modes. We show performance close to the Cram\'er-Rao bounds for these interferometers.

To demonstrate these different interferometers we use a single trapped $^{40}$Ca$^+$ ion as the mechanical oscillator. The ion is confined in a macroscopic, room-temperature, linear Paul trap with an ion-electrode spacing of $\simeq$ 0.75\,mm. Experiments are performed using the two `radial' modes of the ion (normal to the trap axis) with frequencies $\omega_a \approx 2\pi \times 1.80 $\,MHz and $\omega_b \approx 2\pi \times 1.83$\,MHz, and oscillator energy eigenstates denoted by $\ket{n_a,n_b}$. The mode splitting can be adjusted by changing a DC potential on the RF electrodes (see Appendix~\ref{appendix a} for details) which also rotates the mode angle and can be used to vary the achievable coupling rate of the interactions described below. We use qubit states $\ket{\uparrow} \equiv \ket{m_L = + 5/2}$ and $\ket{\downarrow} \equiv\ket{ m_L = +3/2}$ within the metastable $D_{5/2}$ manifold which has a lifetime of $\simeq$ 1.1 seconds~\cite{Kreuter2005}, where $m_L$ is the total angular momentum projection along the direction of the quantization magnetic field of 1.565\,G. The qubit transition frequency is $\omega_0 \approx 2\pi\times 2.63$\,MHz. This is a magnetic field sensitive qubit with a coherence time of $\simeq$ 1 \,ms, while stable trapping potentials provide motional coherence times over 20\,ms. The qubit is manipulated using an orthogonal pair of laser beams red detuned from the transition frequency between $P_{3/2}$ and $D_{5/2}$ manifolds by 44\,THz to induce coherent stimulated-Raman transitions between the two qubit states. In each experiment the qubit is first prepared in the $\ket{\uparrow}$ state with high fidelity, using a sequence of pulses involving 393\,nm, 397\,nm, 854\,nm and 866\,nm lasers. The motion is prepared in $\ket{0,0}$using electromagnetically-induced-transparency (EIT) cooling~\cite{Roos2000} implemented with a pair of 397\,nm beams on the $S_{1/2} \rightarrow P_{1/2}$ transition and resolved-sideband laser cooling using the qubit, which gives the combined initial state of the qubit an motion as approximately $\ket{\uparrow,n_a = 0,n_b = 0}$. True motional mode temperatures correspond to $\bar{n}_{a,b}<0.1$. Qubit readout is performed by ``deshelving" population in the $\ket{\downarrow}$ state using a 854\,nm laser to drive the $D_{5/2} \rightarrow P_{3/2}$ transition followed by applying a laser beam resonant with the $S_{1/2}\leftrightarrow P_{1/2}$ cycling transition and detecting state-dependent ion fluorescence~\cite{Sherman2013}. Motional state analysis is performed by setting the Raman beams difference frequency to $\omega_0 \pm \omega_{m}$, where $m$ denotes the desired mode for readout. Applying these sideband interactions maps the motion onto the qubit states and produces Rabi oscillations with multiple frequency components having amplitudes that depend on the Fock state populations~\cite{Meekhof1996}. This interaction is generated by applying the same pair of Raman beams that drive the qubit, but detuned above (red sideband) or below (blue sideband), the qubit frequency. In the Lamb-Dicke limit~\cite{Meekhof1996}, the red sideband interaction drives transitions between $\ket{\uparrow}\ket{n}$ and $\ket{\downarrow}\ket{n-1}$ with Rabi frequencies proportional to $\sqrt{n}$ and can be used to verify preparation or return to the motional ground state. The blue sideband interaction drives transitions between $\ket{\uparrow}\ket{n}$ and $\ket{\downarrow}\ket{n+1}$ with Rabi frequencies proportional to $\sqrt{n+1}$ and is used to fully characterize the generated state populations~\cite{Meekhof1996}.

\begin{figure}[h]

\centering
    \hspace*{-0.5in}
    \includegraphics[scale = 1.2]{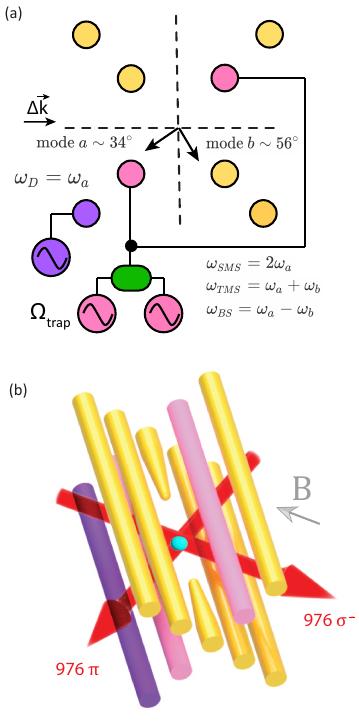}
    \caption{ (a) View of trap along the trap axis showing the electrical connections:\textit{pink}---RF quadrupole electrodes, trap and parametric drive, \textit{purple}---displacement drive and DC electrode, \textit{yellow}--- other DC electrodes and relevant control electrodes. The green box represents the circuitry necessary to combine the drives (Appendix~\ref{appendix a}). Mode $a$ and mode $b$ label the two different `radial' mode vectors of a single ion with the angles from the horizontal shown. (b) A 3D rendering of the trap showing direction of the 976\,nm laser used for motional state preparation and state characterization, and cyan ball representing the $^{40}$Ca$^+$ ion}
\label{fig:1 setup}
\end{figure}

Generation of the motional states is accomplished by applying time-varying potentials to the trap electrodes (see Fig.~\ref{fig:1 setup}) producing a unitary operation. The displacement operation is produced by applying an oscillating potential, at the mode frequency, to a single trap electrode (purple electrode Fig.~\ref{fig:1 setup}). Single-mode squeezing, two-mode squeezing and the beam splitter are accomplished by applying an oscillating potential at twice the mode frequency $(2\omega_a)$, at the sum of the mode frequencies $(\omega_a+\omega_b)$ or at the difference frequency $(\omega_a-\omega_b)$ respectively (pink electrodes Fig.~\ref{fig:1 setup}). These modulated potentials yields a generalized squeezing interaction Hamiltonian,
\begin{equation}
    \hat{H} = i\hbar \dfrac{g}{2}\big[\hat{a}_i \hat{a}_j\exp(i\delta t)\exp(-i\theta)-\hat{a}^\dagger_i \hat{a}^\dagger_j\exp(-i\delta t)\exp(i\theta)\big]
    \label{eq:interactionhamiltonian}
\end{equation}
where $g$ is the parametric coupling strength, which has a different value for each operation, $\delta$ is the detuning from the interaction resonance (either single-mode squeezing (mode $a$ or $b$) or two-mode squeezing), and $\theta$ is the phase of the parametric modulation. With the proper choice of detuning this implements the unitary squeezing operator with squeezing parameter magnitude given by $r_{SMS} = gt$ and the two mode squeezing operator with $r_{TMS} = 2gt$. Here, we used a single-mode displacement coupling of $g = 2\pi\times 1.37$\,kHz which gives a displacement parameter $\alpha = gt$. The maximum single-mode squeezing coupling is $g = 2\pi\times 3.99$\,kHz and a maximum two-mode squeezing coupling of $g = 2\pi\times 1.15$\,kHz. Maximum coupling refers to a coupling rate calculated from fits to the states with the maximum voltage we used and the relevant pulse time for each operation. The beam splitter coupling is $g = 2\pi\times 0.64$\,kHz calculated using calibration data shown in Appendix \ref{appendix c}.

%--------------------------------------------------------------------%

%------------------------------------------------------------------------%

\begin{figure}[h]
\centering
    \includegraphics[scale=1.25]{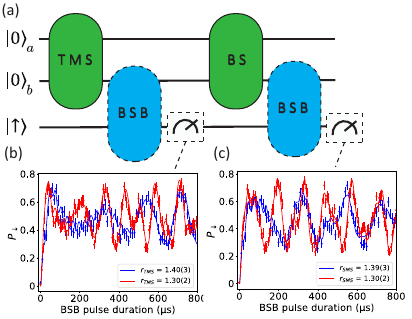}
    \caption{(a) General quantum circuit diagram for characterizing the two-mode squeezed state with blue sideband readout performed at one of two instances, to verify we are producing the expected output (b) Measurement after the two-mode squeezed state and before beam splitter produces a thermal state where the squeezing parameter $r_{TMS}$ is a fit parameter in the model and corresponds to mean occupation $\bar{n}_a= 3.04(3)$ and $\bar{n}_b$ = 3.02(3) for the two measurements. (c) applying the beam splitter operation after the two-mode squeezer and then characterizing the state produces two single-mode squeezed states with squeezing parameter $r_{SMS}$ equal to that of the two-mode squeezed state. The corresponding number distributions for these fits can be found in Fig.~\ref{fig:3 sup. Pn fitting} in Appendix \ref{appendix d}.}
\label{fig:2 TMS verification}
\end{figure}

The beam splitter calibration confirms coherent exchange of phonons between modes even when only reading out on one mode per shot of the experiment. Verification of the two-mode squeezed state is more complicated using the single-mode readout. Using the blue sideband readout method, described above, is equivalent to tracing over a single mode and a thermal state is measured~\cite{Caves1991}. This is analogous to measuring a single qubit from a Bell state which produces a random qubit state. Additionally, we can use a beam splitter operation after the two-mode squeezer to disentangle the modes and produce single-mode squeezed states (Fig.~\ref{fig:2 TMS verification}). The two modes have a different laser coupling, given by their relative orientation, shown in Fig.~\ref{fig:1 setup}a, which gives qualitatively different results for the blue sideband analysis when the motional states are the same. Performing both of these measurements verifies we have produced the target two-mode squeezed state and enables the use of these states for interferometric measurements.

%---------------------------------------------------------------%
%-----------------------------------------------------------------------------------------------------%

\begin{figure}[!htb]
\centering
    \hspace{-20pt}
    \includegraphics[scale = 1.35]{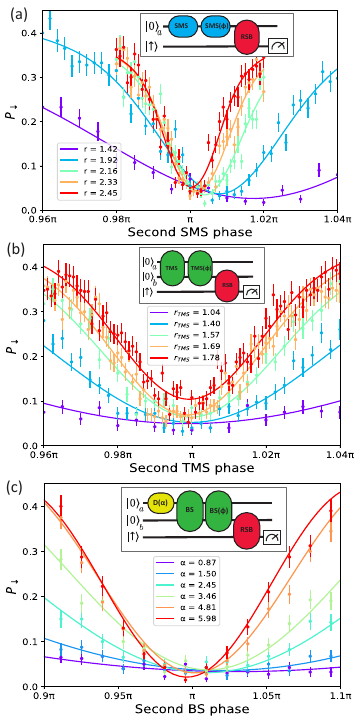}
    \caption{ (a) Single-mode SU(1,1) interferometer output for different amounts of squeezing is shown by varying the relative phase of the second squeezing pulse shown in the circuit diagram (inset). (b) repeat of (a) with two-mode squeezing and two-mode SU(1,1) shown in the interferometer circuit inset.(c) SU(2) interferometer output shown for various sizes of initial displacement where relative phase is controlled with a variable delay shown in the SU(2) circuit inset. All fits are performed using the different analytical expression shown in Appendix \ref{appendix b} where the state parameter $(\alpha,r,r_{TMS})$, vertical and horizontal offsets are fit parameters. }
\label{fig:3 phase scans}
\end{figure}

Implementing time reversal protocols is a method for generating these types of interferometers and is a powerful tool for sensing applications~\cite{Colombo2022}. These protocols also enable us to verify that we are coherently generating the desired states by time-reversing back to the initial state. Starting with a single mode, the SU(1,1) interferometer (Fig.~\ref{fig:3 phase scans}a) is implemented by using a parametric drive tuned to twice the frequency of mode $a$. Adding in the second mode, the SU(2) interferometer (Fig.~\ref{fig:3 phase scans}c) is implemented by first displacing mode $a$ in our experiment, which generates the desired initial coherent state. The beam splitter is then used to generate a superposition of coherent and vacuum states. The two-mode (Fig.~\ref{fig:3 phase scans}b) SU(1,1) interferometers uses the same parametric drive as with the single-mode experiment, but now tuned to the sum frequency of mode $a$ and $b$.

The phase for the single and two-mode SU(1,1) interferometer is varied by programming the relative phase of the second squeezing pulse set by the driving electronics. To avoid increased decoherence, from long pulse durations, these pulses are run at a fixed duration with the drive voltage being varied to control the size of the state. %We report a maximum ideal (no decoherence) single-mode squeezing size of $r = 2.58$ and a ideal two-mode squeezing size of $r = 1.82$ of squeezing.% 
We are limited in the amount of two-mode squeezing we can generate due to our increased sensitivity to common-mode phase fluctuations. 
For the SU(2) interferometer the phase of the final beam-splitter is controlled by using a variable delay, such that the accumulated phase is $(\omega_{a} - \omega_{b})t_{delay}$. The final beam splitter will completely transfer the phonons to mode $b$ with the correct choice of relative phase, and the red sideband readout will verify we have returned to the ground state (Fig.~\ref{fig:3 phase scans}c). We use a maximum displacement of $\alpha \simeq \lvert6\rvert$ to maintain a similar maximum $\bar{n}$ in the single and two-mode squeezed states.

In order to determine the sensitivities to small phase shifts that are achievable with these different experiments, we calculate the quantum Fisher information for a pure state~\cite{Agarwal2022}

\begin{equation}
    \mathcal{F}(\phi) = 4(\Delta \hat{\mathcal{G}})^2 
\end{equation}
where $(\Delta\hat{\mathcal{G}})^2 $ is the variance in the initial state of the probe (coherent state, squeezed state, two-mode squeezed state) and $\hat{\mathcal{G}} = i [d\hat{U}^\dagger/d\phi]\hat{U}$. For phase sensing $\hat{U} = e^{i\phi a^\dagger a}$ and therefore $\hat{\mathcal{G}} = a^\dagger a$ and the Fisher information is proportional to variance in the phonon number of the probe states. The phase sensitivity is limited by the Cram\'er-Rao bound~\cite{Agarwal2022} given by 

\begin{equation}
    \Delta \phi \ge \dfrac{1}{\sqrt{\mathcal{F(\phi)}}}.
\end{equation}
This means for the SU(2) interferometer we are limited at 

\begin{equation}
    \Delta \phi = \dfrac{1}{\alpha} = \dfrac{1}{\sqrt{\expval{N}_d}}
\end{equation}
where $\expval{N}_d$ is the expected mean phonon number for the displacement generated and this limit corresponds to the expected standard quantum limit. The limit for the single-mode SU(1,1) interferometer is given by~\cite{Monras2006}

\begin{equation}
    \Delta \phi = \dfrac{1}{\sqrt{8\expval{N}_{SMS}(\expval{N}_{SMS}+1)}}
\end{equation}
where $ \expval{N}_{SMS}$ is the expected mean phonon number for the input squeezed state. The two-mode SU(1,1) limit is given by~\cite{Agarwal2022}

\begin{equation}
    \Delta \phi = \dfrac{1}{\sqrt{\expval{N}_{TMS}(2+\expval{N}_{TMS})}}
\end{equation}
where $\expval{N}_{TMS}$ is the expected mean phonon number for the input two-mode squeeze state.

The sensitivities achieved are calculated from the experimental data using the Fisher information~\cite{Agarwal2022}
\begin{equation}
    F(\phi) = \sum_j\dfrac{1}{P_j(\phi)}\bigg[\dfrac{dP_j(\phi)}{d\phi}\bigg]^2
\end{equation}
where $P_j(\phi)$ is the probability of getting the experimental results $j$ if the value of the parameter is $\phi$. In our experiment with only two possible outcomes, spin-up or spin-down, this simplifies to 
\begin{equation}
    F(\phi) = \dfrac{1}{P_\downarrow(1-P_{\downarrow})}\bigg[\dfrac{\partial P_\downarrow}{\partial\phi}\bigg]^2.
\end{equation}
%-----------------------------------------------------------------------------------------------------%

%-----------------------------------------------------------------------------------------------------%
The variance and slope of the signal are expressed analytically and are fit to the phase signals that we generate (Fig.~\ref{fig:3 phase scans}). Each interferometer has a unique expression to produce the different fringes shown in Fig.~\ref{fig:3 phase scans}. Knowing the slope and variance of each fitted fringe provides the Fisher information as a function of the phase, where the maximum value of the Fisher information is taken to be the sensitivity shown as the data points in Fig~\ref{fig:4 sensitivities}. These expressions are derived in Appendix \ref{appendix b}, without any adverse effects and represent the ideal experimental output (Fig.~\ref{fig:4 sensitivities}~dashed lines). Calibration of the input state ensures the values of $\bar{n}$ are known without relying on the phase fringe fits. 

Systematic offsets in the phase fringes, seen in the single-mode squeezing phase scans, from $\pi$ are due to an additional pseudopotential~\cite{Leefer2017} when the parametric drive is applied. This additional potential accounts for a maximum shift of $~10\,$kHz of the mode frequencies. In the single-mode SU(1,1) experiment the interaction frequency is calibrated at the maximum coupling meaning as the coupling is turned down, the mode frequencies decrease and larger phase offsets are produced. This phase offset is calibrated away in the two-mode phase scans by calibrating the mode frequencies at every parametric drive voltage. The maximum phase sensitivity is extracted and plotted against the prediction from the quantum Fisher information (Fig.~\ref{fig:4 sensitivities}). The results we achieve deviate from the ideal limit due to motional and qubit decoherence as well as motional heating. Initial mode temperatures also affects these results and contribute to vertical offsets of the fringes in Fig.~\ref{fig:3 phase scans}. The ideal experimental output deviates from the CR bound due to the choice of red side-band readout times, where optimal pulse time varies with $\bar{n}$. Red side-band pulse times were not optimized across all $\bar{n}$ due to the constraints on pulse times due to off-resonant driving of other modes and increased pulse length causing further decoherence.

%-----------------------------------------------------------------------------------------------------%

\begin{figure}[h]
\centering
\hspace{-15pt}
\includegraphics[scale=0.6]{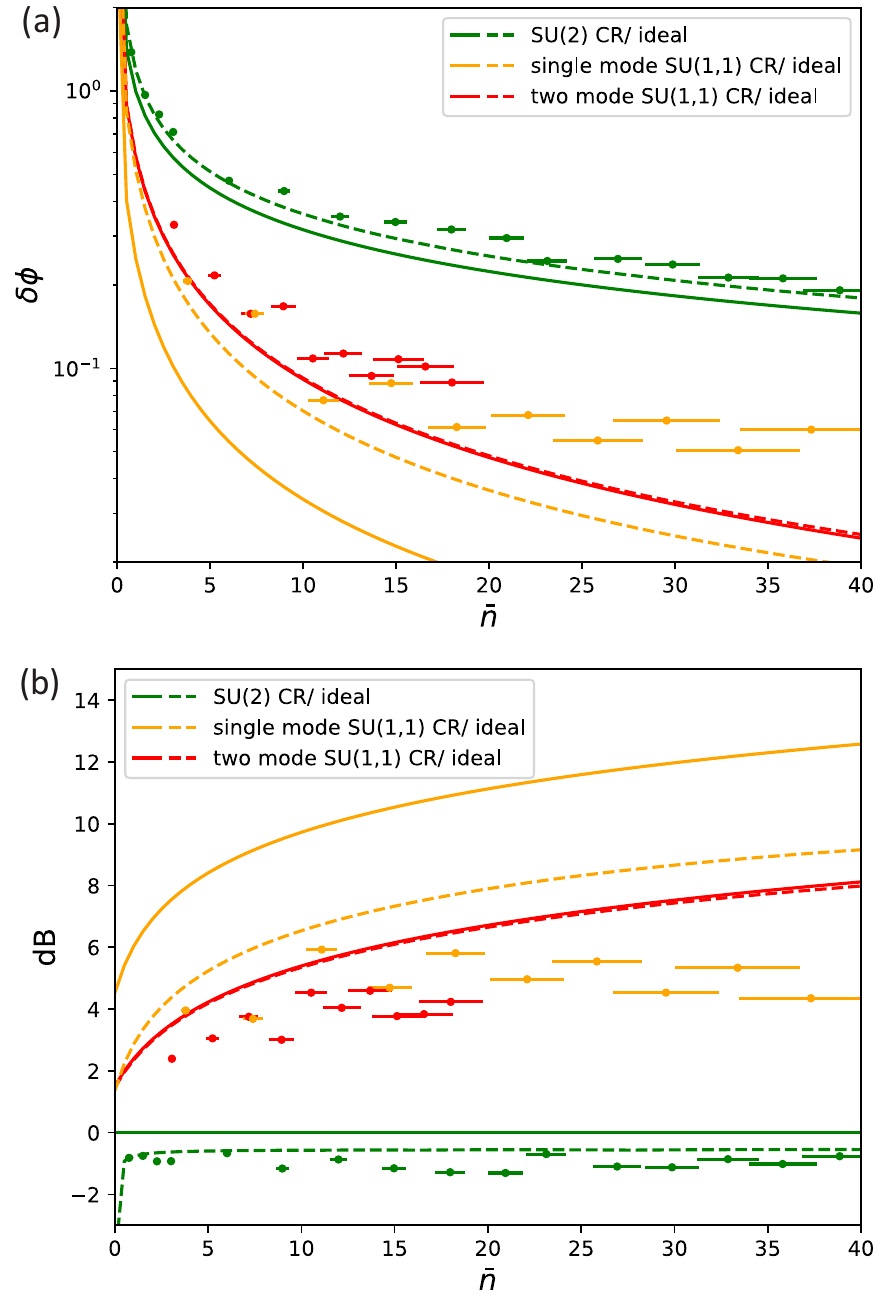}
\vspace{-20pt}
\caption{(a) Ideal experimental phase sensitivities ($\delta\phi$) for the SU(2) (dashed green line), single-mode SU(1,1) (dashed orange line) and two-mode SU(1,1) dashed red line) are shown and can be compared to experimental results shown as dots with error bars and Cram\'er-Rao (CR) bounds in solid lines. The solid green line is the SQL. (b) Sensitivity results for each of the interferometers in terms of dB from the SQL. The 0 dB green line represents the SQL} 

\label{fig:4 sensitivities}
\end{figure}
%---------------------------------------------------------------------------------------------------------------------%

 We have demonstrated a wide class of trapped-ion motional state interferometers using a `laser-free' parametric drive achieving single and two-mode SU(1,1) interferometer sensitivities of 5.9(2)\,dB and 4.5(2)\,dB below the SQL, respectively and a SU(2) interferometer sensitivity within 0.67(5)\,dB of the SQL. These results illustrate the measurement of phase shifts near the Cram\'er-Rao bound for common-mode, relative, and single-mode phase fluctuations. Enhanced sensitivity for measuring these motional phase fluctuations could enable characterization and mitigation of motional frequency noise which can benefit experiments involving motional states. Beside providing a framework for implementing quantum-enhanced sensing protocols, our methods provide the flexibility to concatenate multiple motional operations into circuits for potential applications in CVQC.

 During preparation of this manuscript we became aware of related work using reservoir engineering to prepare two-mode squeezed states of trapped ions~\cite{Li2023}.

\begin{acknowledgements}
   This research is supported by support from NSF through the Q-SEnSE Quantum Leap Challenge Institute, Award \#2016244, and the US Army Research Office under award W911NF-20-1-0037. The data supporting the figures in this article are available upon reasonable request from D.T.C.A. 
\end{acknowledgements}
\appendix
\section{Parametric Drive Electronics}\label{appendix a}

\begin{figure}
    \centering
    \includegraphics[scale=.45]{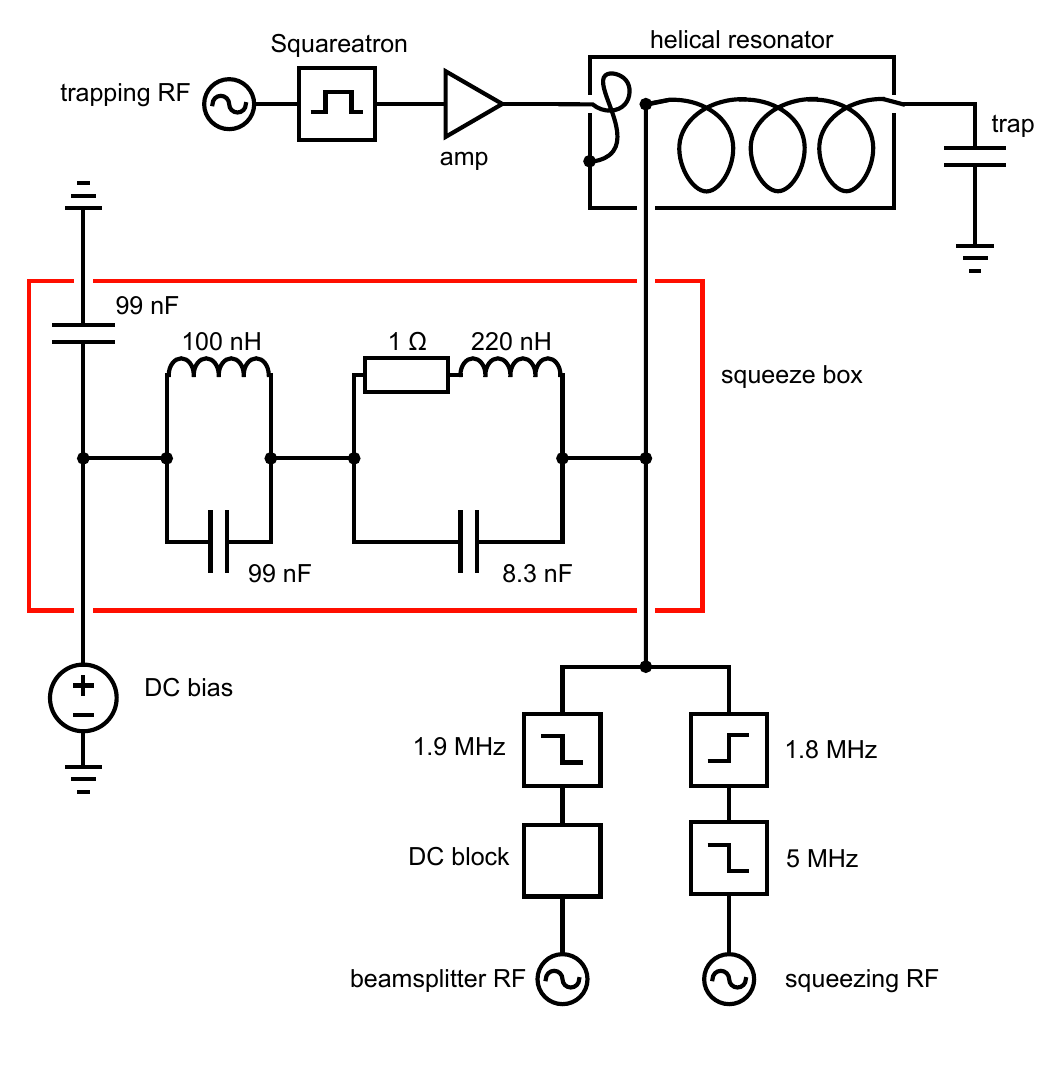}
    \caption{Electrical circuit diagram of the trapping RF, DC bias, and parametric driving RF signals. The beam splitter RF is filtered by a Mini-Circuits SLP-1.9+\,MHz low-pass and a DC block BLK-18W-S+; the squeezing RF is filtered by a Mini-Circuits SLP-5+\,MHz low-pass filter and a custom made 5th order elliptic 1.8\,MHz high-pass filter.}
    \label{fig:squeeze_circuit}
\end{figure}

\begin{figure}
    \centering
    \includegraphics[scale=.47]{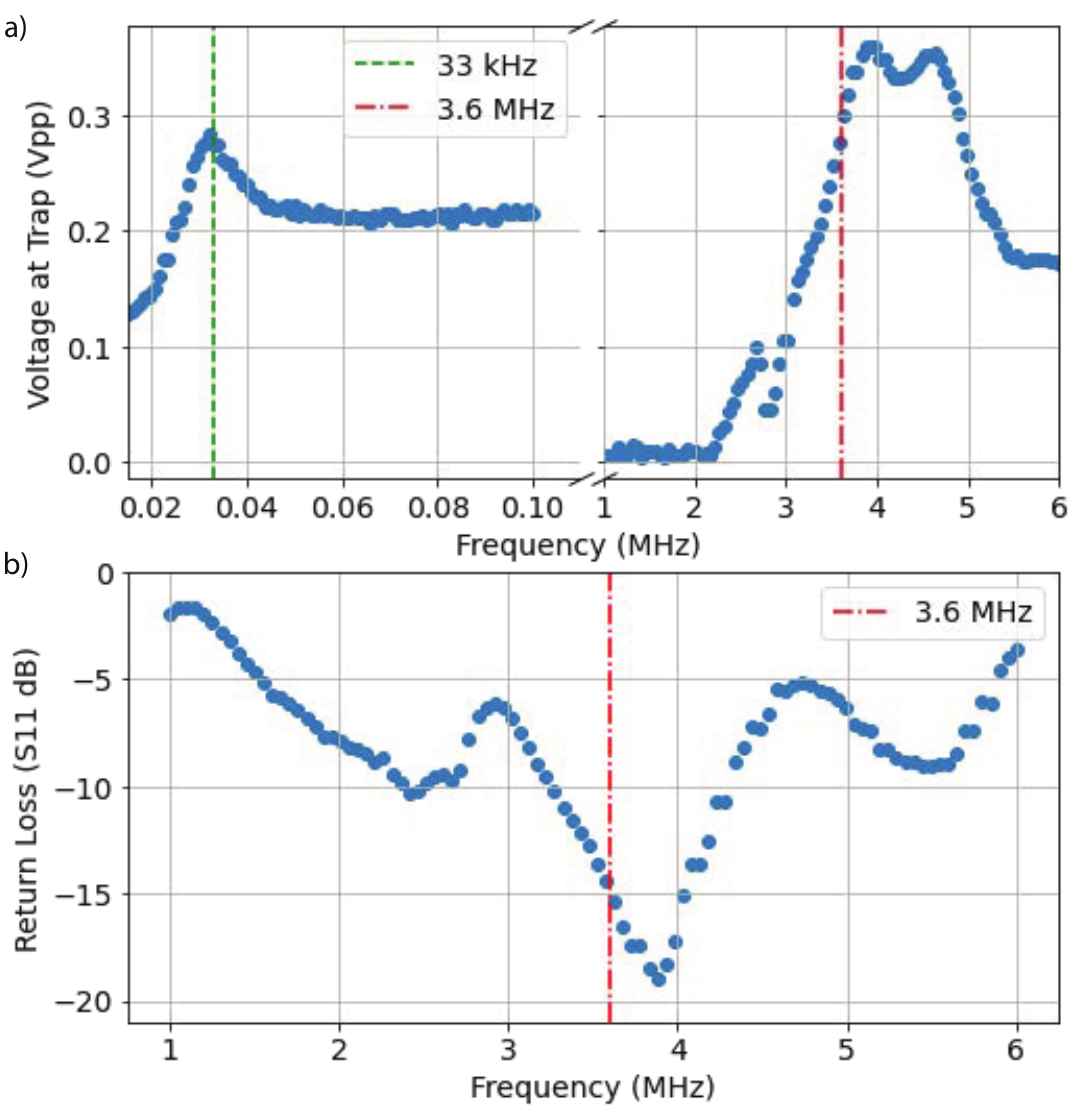}
    \caption{a) Voltage measured at the ion trap vacuum feedthrough with beamsplitter and squeezing RF sources outputting 0\,dBm of power. Vertical dashed green and dash-dot red lines indicate the beam splitter and squeezing frequencies, 33\,kHz and 3.6\,MHz respectively. b) Return loss for the squeezing drive input showing good impedance matching.
}
    \label{fig:squeeze_voltages}
\end{figure}

We apply the voltages required to generate the quadrupole potential for squeezing and beams splitter operations directly to the RF trapping electrodes. This conveniently ensures the quadrupole is automatically well centered on ions located on the axis of our linear trap. The trapping RF is delivered to the trap electrodes by stepping up the 50\,$\Omega$ RF source to high voltage via a helical resonator acting as a transformer, as shown in Fig.~\ref{fig:squeeze_circuit}. This RF drive is amplitude stabilized by the ``Squareatron''~\cite{oregonsquareatron} to improve the motional coherence of the ion's radial modes. The trapping RF and beamsplitter RF are produced by a Rigol DG1022Z arbitrary waveform generator (AWG) and the squeezing RF is produced by an Urukul DDS card~\cite{Kasprowicz2022}, part of our ARTIQ Sinara experimental control system~\cite{Kasprowicz:20}. The parametric drive tones are applied directly to the `back' of the helical resonator's coil via bandpass filters that ensure the two sources are isolated from each other. As these tones are at frequencies well below the helical's resonance they pass through to the trap electrodes relatively unimpeded. Also connected to the back of the resonator coil is the ``squeeze box" containing a pair of LC circuits, which provides three functions. Firstly, it provides a DC bias path to allow control of the mode rotation and splitting. Secondly, it provides a low impedance path to ground at the RF trapping frequency, ensuring the helical resonator works correctly. Compared to a direct ground connection, the ``squeeze box" decreases the quality factor ($Q=f_0/\mathrm{FWHM}$) of the helical modestly, from 310 to 295, and shifts its resonant frequency of 14.5\,MHz by approximately 6\,kHz. Thirdly, it provides an impedance match for the two parametric drive tones by having the LC circuits in the box tuned to match the beamsplitter and squeeze drive frequencies of 33\,kHz and 3.6\,MHz respectively. Electrical testing shows we get voltages on the RF electrodes close to what we expect, as well as low return loss at the squeezing drive frequency Fig.~\ref{fig:squeeze_voltages}.

The potential applied at the trap electrodes determines the parametric coupling strength $g$ for the operations, as shown in Eq.~\ref{eq:interactionhamiltonian}. For measured voltages on the trap RF electrode of 0.26\,V$_{pp}$ for the beam splitter and 0.58\,V$_{pp}$ for both single and two-mode squeezing we get couplings of $g_{BS}=2\pi \cross 0.66\,\mathrm{kHz}$, $g_{SMS}=2\pi \cross 3.68\,\mathrm{kHz}$ and $g_{TMS}=2\pi \cross 1.09\,\mathrm{kHz}$ respectively. These coupling values are calculated from theory using the above voltages and agree reasonably with the values calculated from experiment shown in the main text.

%--------------------------------------------------------------------------------------------------------------%
%-------------------------------------------------------------------------------------------------------------%
%-------------------------------------------------------------------------------------------------------------%

\section{Experimental output and Fisher information}\label{appendix b}
To obtain an analytical expression for the output of these interferometers we must determine the action of the red sideband on the final state and determine the expectation values for the final qubit projection operator. We calculate this below for the two-mode SU(1,1) case. The single-mode case can be treated in a similar manner. We start by representing the two-mode squeezed state in the Fock basis~\cite{Kurochkin2014}

\begin{equation}
\begin{aligned}
    \ket{TMS} = \dfrac{1}{\cosh{r}}\sum_{n = 0}^{\infty} \tanh^n{r}\ket{n,n} &= \\
    \sqrt{\dfrac{1}{1+\lambda}}\sum_{n=0}^{\infty}\left(\sqrt{\dfrac{\lambda}{1+\lambda}}\right)^n\ket{n,n}
\end{aligned}
\end{equation}
where $\lambda = \sinh{r}$. We need to determine $\ket{out} = TMS(\phi)TMS\ket{0,0}$. For phase other than $\phi = \pi$ the output is still a two-mode squeezed state with a different two-mode squeeze parameter $r$, which we can express as a function of the phase.

\begin{equation}
    r(\phi) = \sinh^{-1}({\sinh{r_0}\cos{\phi/2}}),
\end{equation}
where $r_0$ is the input two-mode squeeze state magnitude. The red sideband Hamiltonian is written as~\cite{Meekhof1996}

\begin{equation}
    H_{RSB} = \hbar\eta\dfrac{\Omega}{2}(\sigma^+a+\sigma^-a^\dagger).
\end{equation}

$\Omega$ is the qubit carrier Rabi frequency, $\eta$ is the Lamb-Dicke parameter~\cite{Meekhof1996} for the given mode, and $\sigma^+$ and $\sigma^-$ are the single qubit raising and lowering operators. The unitary operator for the Hamiltonian can be written as 

\begin{equation}
    U = e^{\dfrac{i\beta H_{RSB}t}{\hbar}}
\end{equation}
where $\beta = \eta \Omega t/2$. Expanding this unitary into an infinite series gives

\begin{equation}
    \ket{n,n}\rightarrow \sum_{j = 0}^{\infty} \dfrac{(i\beta)^j}{j!}(\sigma^+a+\sigma^-a^\dagger)^j\ket{\downarrow,n,n}.
\end{equation}

Only terms having equal numbers of raising and lower operators for spin and motion will have a final contribution to the projection, so we can re-write this state

\begin{equation}
    U_{RSB}\ket{\downarrow,n,n} = \sum_{j = 0}^{\infty} \dfrac{(i\beta)^j}{j!}(\sigma^+a\sigma^-a^\dagger)^j\ket{\downarrow,n,n},
\end{equation}
where $\{j\}$ is only even integers. This leaves the final spin state unaffected and only the number operator $a^\dagger a$ remains, leaving us

\begin{equation}
\begin{aligned}
    U_{RSB}\ket{\downarrow,n,n} =  \sum_{j = 0}^{\infty}\dfrac{(i\beta)^{2j}}{(2j)!}n^j\ket{\downarrow,n,n} &= \\
     \sum_{j = 0}^{\infty}\dfrac{(i\beta)^{2j}}{(2j)!}\sqrt{n}^{2j}\ket{\downarrow,n,n} = \cos{(\beta\sqrt{n})}\ket{\downarrow,n,n}.
\end{aligned}
\end{equation}

The expectation value of the spin projector is therefore

\begin{equation}
\begin{aligned}
    \ev{U_{RSB}^\dagger\ket{\downarrow\otimes \identity}\bra{\downarrow\otimes\identity}U_{RSB}}{\downarrow,TMS} &=\\
    \dfrac{1}{1+\lambda}\sum_{n = 0}^{\infty}\cos{(\sqrt{n}\eta\Omega\dfrac{t}{2})\bigg(\dfrac{\lambda}{1+\lambda}\bigg)^n}
\end{aligned}
\end{equation}
For the sake of simplicity we will call this expectation value $\expval{\hat{X}}$ and we define the variance
\begin{equation}
    \Delta\hat{X}^2 = \expval{\hat{X}^2}-\expval{\hat{X}}^2.
\end{equation}
For spin projection $\expval{\hat{X}^2} = \expval{\hat{X}}$. We also need an expression for the derivative of the expectation value as a function of the phase;

\begin{equation}
\begin{gathered}
    \dfrac{d\expval{\hat{X}}}{d\phi} = \dfrac{d\lambda}{d\phi}\dfrac{1}{1+\lambda}  \\ \bigg(\expval{\hat{X}}+\dfrac{1} {1+\lambda}\sum_{n=0}^{\infty}\cos^2{(\sqrt{n}\eta\dfrac{\Omega t}{2})n\bigg(\dfrac{\lambda}{1+\lambda}\bigg)^{n-1}}\\
    \bigg(1-\dfrac{\lambda}{1+\lambda}\bigg)\bigg)
\end{gathered}
\end{equation}
where $\dfrac{d\lambda}{d\phi} = -2\cosh^2{r}\sin{\phi}\sinh^2{r}$. The same process is carried out and shown for the coherent state in the SU(2) interferometer as these results differ significantly from the two-mode SU(1,1). A coherent state is expressed in the Fock basis as

\begin{equation}
    \ket{\alpha} = e^{-\abs{\alpha}^2/2}\sum_{n=0}^{\infty}\dfrac{\alpha^{2n}}{n!}\ket{n}.
\end{equation}
The red sideband produces a similar output,

\begin{equation}
\begin{aligned}
    \ev{U_{RSB}^\dagger\ket{\downarrow\otimes \identity}\bra{\downarrow\otimes\identity}U_{RSB}}{\downarrow,\alpha} = \expval{\hat{X}} =  \\
    e^{-\abs{\alpha}^2}\sum_{n=0}^{\infty}\dfrac{\alpha^{2n}}{n!}\cos^2{(\sqrt{n}\eta\Omega t/2)}
\end{aligned}
\end{equation}
with $\alpha$ a function of the phase, $\alpha = \alpha_0\sin{\phi/2}$. The variance can again be expressed in terms of this expectation value as in (17) and the derivative of $\langle\hat{X}\rangle$ is given as

\begin{equation}
    \dfrac{d\expval{\hat{X}}}{d\phi} = e^{\abs{\alpha}^2}\sum_{n=0}^{\infty}\dfrac{\alpha^{2n}}{n!}\cos^2{(\sqrt{n}\eta\Omega t/2)}(-\alpha^{2n+1}+n\alpha^{2n-1})
\end{equation}

\section{Beam splitter calibration}\label{appendix c}

\begin{figure}[h]
\centering
    \includegraphics[scale=0.6]{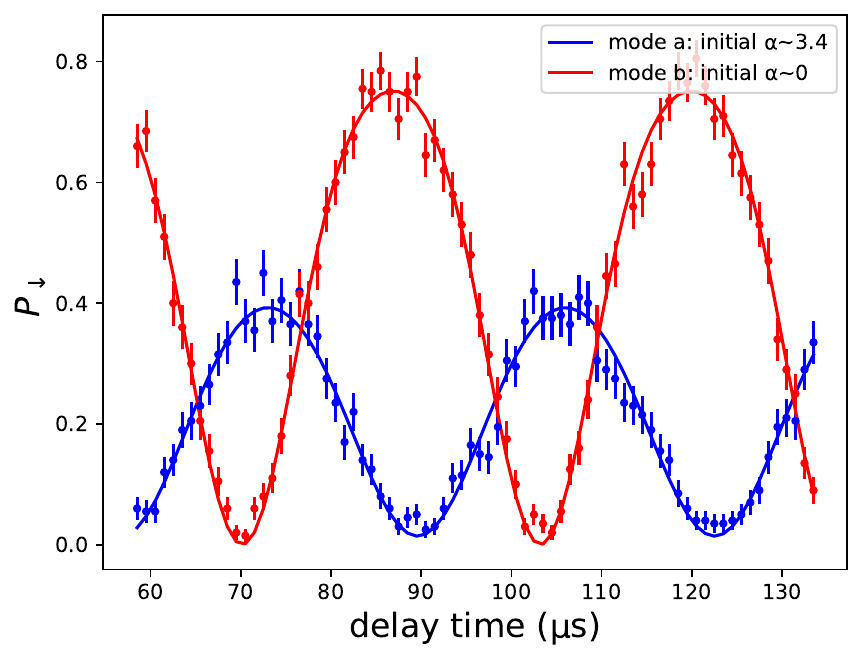}
    \caption{Starting with a coherent state in the low-frequency mode, the SU(2) interferometer is performed with a variable delay between beam splitter pulses, allowing coherent transfer of the coherent state between modes. The amplitude of the 50/50 beam splitter pulses used can be calibrated by maximizing contrast on mode $a$ fringes. Delay time is time passed between beam splitter pulses.}
\label{fig:2 sup. BS calibration}
\end{figure}

To produce the beam splitter operation described by the Hamiltonian~\cite{Leibfried2002}
\begin{equation}
    \hat{H} = i\hbar \dfrac{g}{2}\big[\hat{a} \hat{b}^{\dagger}\exp(-i\theta)-\hat{a}^\dagger \hat{b}\exp(i\theta)\big],
\end{equation} 
we digitally trigger the drive AWG to output a pulse consisting of an integer number of cycles (30 cycles in the runs in Fig.~\ref{fig:2 sup. BS calibration}) at a fixed amplitude, starting at a phase of zero. We cannot continuously adjust the length of these pulses and therefore must calibrate the pulse amplitude to produce a 50/50 beam splitter pulse. Specifically, we select an amplitude that maximizes the fringe contrast of our SU(2) interferometer (i.e. that allows a pair of 50/50 beam splitter pulses to fully transfer a displacement from one mode to another). The starting phase of each pulse is fixed which means to adjust the relative phase between beam splitter pulses in our SU(2) experiments we adjust the delay time $t_{delay}$ between pulses, giving a phase offset of $(\omega_{a} - \omega_{b})t_{delay}$, as shown in Fig.~\ref{fig:2 sup. BS calibration}.
%-----------------------------------------------------%
%-----------------------------------------------------%
\section{Squeezed state fitting}\label{appendix d}

For performing motional state analysis we tune our Raman beams resonantly to a blue motional sideband (BSB). This interaction ideally realizes the anti Jaynes-Cummings (JC) Hamiltonian
\begin{equation}
    \hat{H}_{JC} = \hbar \Omega/2(\hat{\sigma}^{+}\hat{a}^{\dagger}+ \hat{\sigma}^- \hat{a})
\end{equation}
where the qubit state transition operators $\hat{\sigma}^{\pm} \equiv \dfrac{1}{2}(\hat{\sigma}_x\mp i \hat{\sigma}_y) $ are constructed from the Pauli operators $\hat{\sigma}_x$ and $\hat{\sigma}_y$ and $\hat{a}(\hat{a}^\dagger)$ are the motional state annihilation (creation) operators. This allows for an analytical solution for the qubit state probabilities to be written\cite{Meekhof1996}
\begin{equation}
    P_\downarrow(t) = \dfrac{1}{2}\Bigg[1+\sum_{n=0}^\infty P(n) \cos{(\Omega_{SB}\sqrt{n+1}t})\bigg]
    \label{eq:interactionhamiltonian_offres}
\end{equation}
where $n$ is the oscillator Fock state, $t$ is the duration of the sideband integration and $\Omega_{SB} = \eta\Omega_{carrier}$ is the sideband Rabi frequency. This enables a general fit of the Fock state probabilities which assumes no ideal model state. 
%----------------------------------------------------------------------------------------------------%
\begin{figure}[h]
\centering
    \includegraphics[scale=1.0]{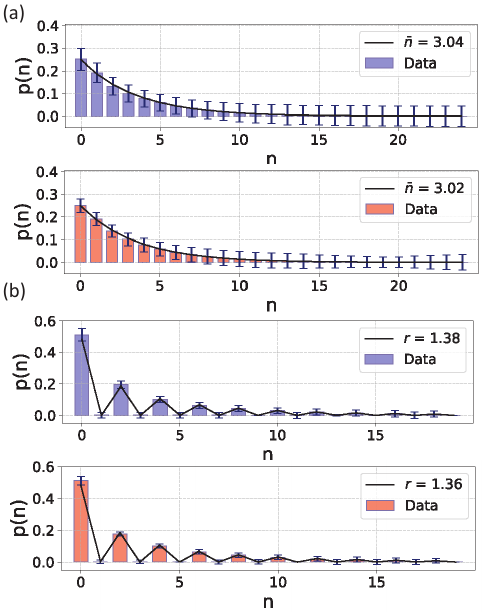}
    \caption{(a) Fits to the output of two-mode squeezing under interaction described above. Blue bars indicate mode $a$ and red bars are mode $b$ with associated error bars. Black line is a thermal state model fit to the Fock state fits. (b) Fock state fits to output of the beam splitter after two-mode squeezing. Blue data is mode $a$ and red is mode $b$ with associated error bars. Black line is a squeezed state model fit to the Fock state fits. }
\label{fig:3 sup. Pn fitting}
\end{figure}
%---------------------------------------------------------------------------------------------------------%
Due to the narrow frequency splitting of our motional modes, the sideband interaction results in off-resonant driving of our second mode which is separated in frequency from the first mode by only $33$\,kHz. The resulting interaction Hamiltonian is now more accurately written as
\begin{equation}
    \hat{H} = \hbar \Omega/2(\hat{\sigma}^{+}\hat{a}^{\dagger}e^{i\delta_1t}+ \hat{\sigma}^- \hat{a}e^{-i\delta_1t}+\hat{\sigma}^{+}\hat{b}^{\dagger}e^{i\delta_2t}+ \hat{\sigma}^- \hat{b}e^{-i\delta_2t})
\end{equation}
where $\delta_1 - \delta_2\approx 2\pi \times33$\,kHz. There is no known analytical solution for the qubit probabilities under evolution of this Hamiltonian therefore (\ref{eq:interactionhamiltonian_offres}) cannot be used. In order to get a model free fit, master equation simulations were performed using QuTiP~\cite{Qutip} where each Fock state probability in the inital state is a fit parameter while preserving the trace. Estimates of fit uncertainties with this method proved the be a challenge due to weak dependence on the off-resonant mode. Fitting was performed starting with an ideal state and holding all Fock states except one constant. The single Fock state was iterated through all Fock states for both mode $a$ and $b$ which may produce an under estimate of the uncertainty. Data where either mode $a$ or $b$ was driven resonantly was fit independently. Figure~\ref{fig:3 sup. Pn fitting} shows the results of the fit Fock states to the data shown in Fig.~\ref{fig:2 TMS verification}, with the model state fit to the probabilities for both modes independently and overlaid. The results of this fitting procedure return approximately the same state as the fit in Fig.~\ref{fig:2 TMS verification} which provides further confirmation of the state that is being generated.
\bibliography{phase_sense_refs.bib}

%\pagebreak
%\section{Referee response}
%\input{ref_B_response}

\end{document}